\documentclass{revtex4}

\input epsf
\usepackage{amsmath,amssymb}
\usepackage{graphicx}

\draft

\begin{document}
\titlepage
\title{Cosmological Evolution in 1/R-Gravity Theory}
\author{Xinhe Meng$^{1,2}$ \footnote{mengxh@public.tpt.tj.cn}
 \ \ Peng Wang$^1$ \footnote{pewang@eyou.com}
} \affiliation{1.  Department of Physics, Nankai University,
Tianjin 300071, P.R.China \\2. Institute of Theoretical Physics,
CAS, Beijing 100080, P.R.China}

\begin{abstract}
Recently, corrections of the $L(R)$ type to Einstein-Hilbert
action that become important at small curvature are proposed.
Those type of models intend to explain the observed cosmic
acceleration without dark energy. We derive the full Modified
Friedmann equation in the Palatini formulation of those modified
gravity model of the $L(R)$ type. Then, we discuss various
cosmological predictions of the Modified Friedmann equation.
\end{abstract}

\maketitle

\textbf{1. Introduction}

It now seems well-established that the expansion of our universe
is currently in an accelerating phase. The most direct evidence
for this is from the measurements of type Ia supernova
\cite{Perlmutter}. Other indirect evidences such as the
observations of CMB by the WMAP satellite \cite{Spergel},
large-scale galaxy surveys by 2dF and SDSS \cite{hui} also seem
supporting this.

But now the mechanisms responsible for this acceleration are not
very clear. Many authors introduce a mysterious cosmic fluid
called dark energy to explain this (see Ref.\cite{Peebles} for a
review). On the other hand, some authors suggest that maybe there
does not exist such mysterious dark energy, but the observed
cosmic acceleration is a signal of our first real lack of
understanding of gravitational physics \cite{Lue}. An example is
the braneworld theory of Dvali et al. \cite{Dvali}. Recently,
there are active discussions in this direction by modifying the
action for gravity [7-17]. Specifically, a $1/R$ term is suggested
to be added to the action \cite{Carroll}: the so called $1/R$
gravity. It is interesting that such term may be predicted by
string/M-theory \cite{Odintsov-string}. In Ref.\cite{Vollick},
Vollick used Palatini variational principle to derive the field
equations for $1/R$ gravity. In Ref.\cite{Dolgov}, Dolgov et al.
argued that the fourth order field equations following from the
metric variation suffer serious instability problem. If this is
indeed the case, the Palatini formulation appears quite appealing,
because the second order field equations following from Palatini
variation are free of this sort of instability \cite{Meng}. In our
previous paper \cite{Meng}, we have derived the first and second
order Modified Friedmann (MF) equations by directly computing the
various order Ricci tensors. We have shown that the first and
second order MF equation can fit the SNe Ia data at an acceptable
level. However, for cosmological predictions involving the
derivative of the Hubble parameter such as the deceleration
parameter, there were obvious differences in the first order and
second order MF equations. This indicted that to make predictions
about those quantities, the approximated Friedmann equation is not
suitable. In this paper, we will derive the full Modified
Friedmann equation for Palatini formulation of the $1/R$ gravity.
We will show that the full Modified Friedmann equation can fit the
SNe Ia data at an acceptable level, which is not surprising from
our previous study. But now we can compute exactly the quantities
involving the derivatives of the Hubble parameter such as the
deceleration parameter and effective equation of state. In this
paper, when computing various cosmological predictions, we will
always use $H_0=70$ km/(sec$\cdot$Mpc) as the current value of the
Hubble parameter.

\textbf{2. The Modified Friedmann Equation}

The field equations follow from the variation in Palatini approach
of the generalized Einstein-Hilbert action (See Ref.\cite{Vollick}
for details)
\begin{equation}
S=-\frac{1}{2\kappa}\int{d^4x\sqrt{-g}L(R)}+\int{d^4x\sqrt{-g}L_M}\label{action}
\end{equation}
where $\kappa =8\pi G$, $L$ is a function of the scalar curvature
$R$ and $L_M$ is the Lagrangian density for matter. In the
Palatini formulation, the connection is not associated with
$g_{\mu\nu}$, but with $h_{\mu\nu}\equiv L'(R)g_{\mu\nu}$, which
is known from varying the action with respect to $\Gamma
^{\lambda}_{\mu\nu}$.

The field equations in Palatini formulation are
\begin{equation}
L'(R)R_{\mu\nu}-\frac{1}{2}L(R)g_{\mu\nu}=-\kappa
T_{\mu\nu}\label{2.2}
\end{equation}
\begin{equation}
R_{\mu\nu}=R_{\mu\nu}(g)-\frac{3}{2}(L')^{-2}\nabla _{\mu}L'\nabla
_{\nu}L' +(L')^{-1}\nabla _{\mu}\nabla
_{\nu}L'+\frac{1}{2}(L')^{-1}g_{\mu\nu}\nabla _{\sigma}\nabla
^{\sigma}L'\label{Ricci}
\end{equation}
\begin{equation}
R=R(g)+3(L')^{-1}\nabla _{\mu}\nabla ^{\mu}
L'-\frac{3}{2}(L')^{-2}\nabla_{\mu}L'\nabla^{\mu}L'\label{scalar}
\end{equation}
where a prime denotes differentiation with respect to $R$,
$R_{\mu\nu}(g)$ is the Ricci tensor with respect to $g_{\mu\nu}$
and $R=g^{\mu\nu}R_{\mu\nu}$, $T_{\mu\nu}$ is the energy-momentum
tensor given by
\begin{equation}
T_{\mu\nu}=-\frac{2}{\sqrt{-g}}\frac{\delta S_M}{\delta
g^{\mu\nu}}\label{2.3}
\end{equation}
We assume that the universe contains dust and radiation, thus
$T^{\mu}_{\nu}=\{-\rho_m-\rho_r,p_r,p_r,p_r\}$ where $\rho_m$ and
$\rho_r$ are the energy densities for dust and radiation
respectively, $p_r$ is the pressure of the radiation. Note that
$T=g^{\mu\nu}T_{\mu\nu}=-\rho_m$ because of the relation
$p_r=\rho_r/3$.

Note by contracting Eq.(\ref{2.2}), we get:
\begin{equation}
L'(R)R-2L(R)=-\kappa T\label{R(T)}
\end{equation}
Assume we can solve $R$ as a function of $T$ from Eq.(\ref{R(T)}).
Thus Eqs.(\ref{Ricci}), (\ref{scalar}) do define the Ricci tensor
with respect to $h_{\mu\nu}$.

Now consider the Robertson-Walker metric describing the
cosmological evolution,
\begin{equation}
ds^2=-dt^2+a(t)^2(dx^2+dy^2+dz^2)\label{metric}
\end{equation}
We only consider a flat metric, which is favored by present
observations \cite{Spergel}.

From Eqs.(\ref{metric}), (\ref{Ricci}), we can get the
non-vanishing components of the Ricci tensor:
\begin{equation}
R_{00}=3\frac{\ddot{a}}{a}-\frac{3}{2}(L')^{-2}(\partial_0{L'})^2+\frac{3}{2}(L')^{-1}\nabla_0\nabla_0L'\label{R00}
\end{equation}
\begin{eqnarray}
R_{ij}=-[a\ddot{a}+2\dot{a}^2+(L')^{-1}\{^0_{ij}\}_g\partial_0L'
+\frac{a^2}{2}(L')^{-1}\nabla_0\nabla_0L']\delta_{ij}\label{ij}
\end{eqnarray}

Substituting equations (\ref{R00}) and (\ref{ij}) into the field
equations (\ref{2.2}), we can get
\begin{equation}
6H^2+3H(L')^{-1}\partial_0L'+\frac{3}{2}(L')^{-2}(\partial_0L')^2=\frac{\kappa
(\rho+3p)+L}{L'}\label{aa}
\end{equation}
where $H\equiv \dot{a}/a$ is the Hubble parameter, $\rho$ and $p$
are the total energy density and total pressure respectively. This
is the general form of the MF equation for Palatini formulation of
the modified gravity of the $L(R)$ type. Assume that we can solve
$R$ in term of $T$ from Eq.(\ref{R(T)}) for a specifical model,
substitute it into the expressions for $L'$ and $\partial_0L'$, we
can get the MF equation.

In this paper we will study the special $L(R)$ suggested in
Ref.\cite{Carroll}, the so called $1/R$ gravity:
\begin{equation}
L(R)=R-\frac{\alpha ^2}{3R}\label{L(R)}
\end{equation}
where $\alpha$ is a positive constant with the same dimension as
$R$ and following Ref.\cite{Vollick}, the factor of $1/3$ is
introduced to simplify the field equations.

The field equations follow from Eq.(\ref{2.2})
\begin{equation}
(1+\frac{\alpha
^2}{3R^2})R_{\mu\nu}-\frac{1}{2}g_{\mu\nu}(R-\frac{\alpha
^2}{3R})=-\kappa T_{\mu\nu}\label{field equ}
\end{equation}

Contracting the indices gives
\begin{equation}
R=\frac{1}{2}\alpha [\frac{\kappa
T}{\alpha}-2\sqrt{1+\frac{1}{4}(\frac{\kappa
T}{\alpha})^2}]\label{R}
\end{equation}

From Eqs.(\ref{field equ}), (\ref{R}) we can see that the field
equations reduce to the Einstein equations if $|\kappa T| \gg
\alpha$. On the other hand, when $\alpha \gg |\kappa T|$,
deviations from the Einstein's theory will be large. This is
exactly the case we are interested in. We hope it can explain
today's cosmic acceleration.

From the conservation equation $\dot{\rho}+3H\rho=0$, we can get
\begin{equation}
\partial_0L'=\frac{(\frac{\alpha}{R})^2\frac{\kappa \rho}{\alpha}}{\sqrt{1+\frac{1}{4}(\frac{\kappa \rho}{\alpha})^2}}H
\label{}
\end{equation}

Substituting this into Eq.(\ref{aa}), we can get the Modified
Friedmann (MF) equation in this modified gravity theory:
\begin{equation}
H^2=\frac{\kappa \rho-\alpha(G(\frac{\kappa
\rho}{\alpha})-\frac{1}{3G(\frac{\kappa
\rho}{\alpha})})}{(1+\frac{1}{3G(\frac{\kappa
\rho}{\alpha})^2})(6+3F(\frac{\kappa
\rho}{\alpha})[1+\frac{1}{2}F(\frac{\kappa
\rho}{\alpha})])}\label{MF}
\end{equation}
where the two functions $G$ and $F$ are defined as
\begin{equation}
G(x)=-(\frac{1}{2}x+\sqrt{1+\frac{1}{4}x^2})\label{G}
\end{equation}
\begin{equation}
F(x)=\frac{x}{(G(x)^2+\frac{1}{3})\sqrt{1+\frac{1}{4}x^2}}\label{F}
\end{equation}
From Eqs.(\ref{G}), (\ref{F}) we can see that when $x\gg 1,
G(x)\sim -x, F(x)\sim 0$. Thus when $\kappa\rho\gg\alpha$,
Eq.(\ref{MF}) reduces to the standard Friedmann equation
\begin{equation}
H^2=\frac{\kappa}{3}\rho\label{}
\end{equation}
This confirms the assertion we made below equation (\ref{R}).
Below we will use the full Modified Friedmann equation (\ref{MF})
to compute numerically some cosmological parameters, namely, age
of the universe in Sec.3, redshift-luminosity relationship in
Sec.4, deceleration parameter in Sec.5, effective equation of
state in Sec.6.

\textbf{3. Age of the universe}

\begin{figure}
  \includegraphics[width=0.5\columnwidth]{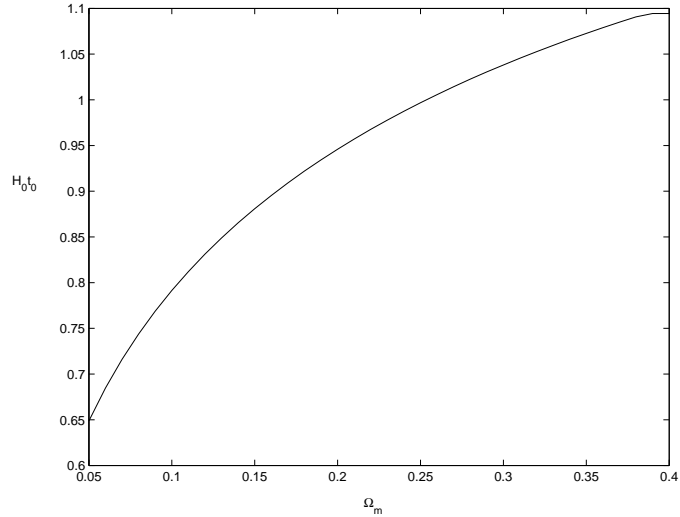}
  \caption{The dependence of the age of the universe on $\Omega_m$.}\label{1}
\end{figure}

The age of the universe is given by
\begin{equation}
t_0=\int_0^1\frac{da}{aH(a)}\label{age}
\end{equation}

Thus from the MF equation (\ref{MF}) we can numerically draw the
dependence of the age of the universe on $\Omega_m$, see Fig.1. If
we take $t_0>10$ Gyr, this implies $H_0t_0>0.71$; if we take
$t_0>11$ Gyr \cite{Krauss}, this implies $H_0t_0>0.79$. From Fig.1
this correspond to $\Omega_m>0.07$ and $\Omega_m>0.1$
respectively.

\textbf{4. Data Fitting with SNe Ia Observations}

\begin{figure}
  \includegraphics[width=0.5\columnwidth]{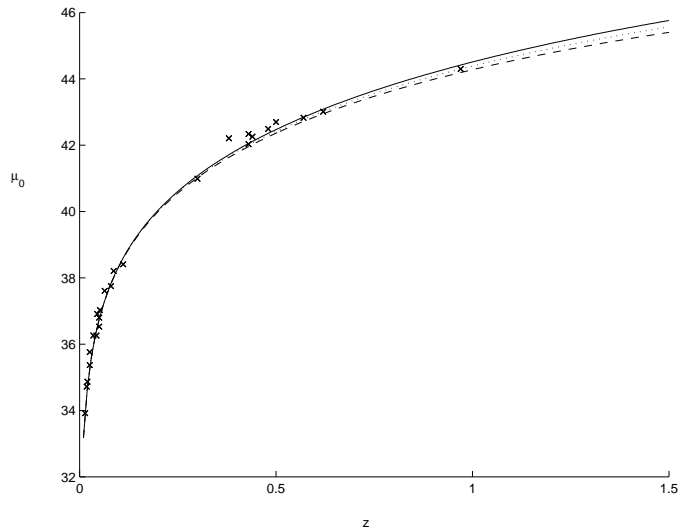}
  \caption{The dependence of luminosity distance on redshift computed from the
  MF equation (\ref{MF}). The solid, dotted, dashed lines correspond to $\Omega_m=0.1, 0.2, 0.3$, respectively. The little crosses
  are the observed data }\label{1}
\end{figure}

\begin{figure}
  \includegraphics[width=0.5\columnwidth]{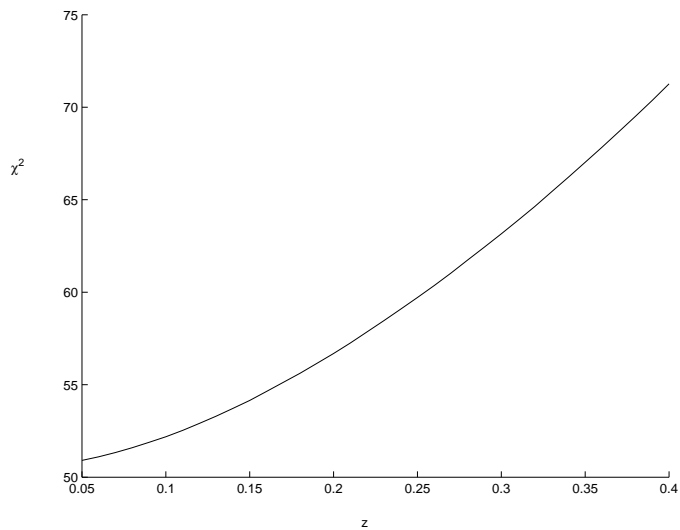}
  \caption{The dependence of the $\chi^2$ on the parameter $\Omega_m$. It can be seen that $\chi^2$ gets a little smaller for
smaller value of $\Omega_m$.}\label{1}
\end{figure}

It is the observations of the SNe Ia that first reveal our
universe is in an accelerating phase. It is still the most
important evidence for acceleration and the best discriminator
between different models to explain the acceleration. Thus, any
model attempting to explain the acceleration should fit the SNe Ia
data as the basic requirement.

We have shown in Ref.\cite{Meng} that the first order MF equation
fits the SNe Ia data at an acceptable level and the second order
correction to the luminosity distance is small, thus we have
concluded that this strongly implies that the full MF equation
also fits the SNe Ia at an acceptable level. Now with the full MF
equation in hand, we can explicitly show this is true and it can
be checked numerically that the deviations between the full MF
equation and the approximated MF equation is very small.

We will use data listed in Ref.\cite{Fabris}, which contains 25
SNe Ia observations. Since our purpose is to show the MF equations
fit the data at an acceptable level, also because the differences
in various $\Omega_m$ get larger when redshift is larger and
trustable observations around redshift 1 is very few, we think 25
low redshift samples is enough. Also we did not perform a detailed
$\chi ^2$ analysis, this is suitable when we get more high
redshift samples.

Fig.2 shows the prediction of the full MF equation for
$\Omega_m=0.1, 0.2, 0.3$ respectively.

Fig.3 shows the dependence of the $\chi^2$ (See Ref.\cite{Meng}
for the definition of this quantity) on the $\Omega_m$. We can see
that $\chi^2$ gets smaller for smaller value of $\Omega_m$. This
property has already been observed when data fitting with the
first order MF equation \cite{Meng}. We think this is an
interesting feature of this modified gravity theory: It has
already provide the possibility of eliminating the necessity of
dark energy for explanation of cosmic acceleration, now it seems
that a universe without dark matter is favored when fit this model
to SNe Ia data. Thus we boldly suggest that maybe this modified
gravity theory can provide the possibility of eliminating dark
matter. This is surely an interesting thing and worth some
investigations.

However, we should also note that we only use SNe Ia data smaller
than redshift 1 to draw the above conclusions. It can easily been
seen in Fig.2 that the differences between predictions drawn from
the different values of $\Omega_m$ become larger when redshift is
around 1. Thus, future high redshift supernova observations may
give a more conclusive discrimination between the parameters.

\textbf{5. The Deceleration Parameter}

Given the observation that our universe is currently expanding in
an accelerating phase, the deceleration parameter should become
negative in recent cosmological times. The deceleration parameter
is defined by $q=-\ddot{a}/aH^2$.

The deceleration parameter for the MF equation (\ref{MF}) is too
complicated and thus we will not present its analytic expression
here. We just draw it in in Fig.4. In Fig.4 it can be also noted
that the redshift $z_{q=0}$ which gives $q(z_{q=0})=0$ gets
smaller for larger value of $\Omega_m$. Current observation of SNe
Ia gives a constraint: $0.6<z_{q=0}<1.7$ \cite{Perlmutter}. Thus
$\Omega_m$ can not be too small.

\begin{figure}
  \includegraphics[width=0.5\columnwidth]{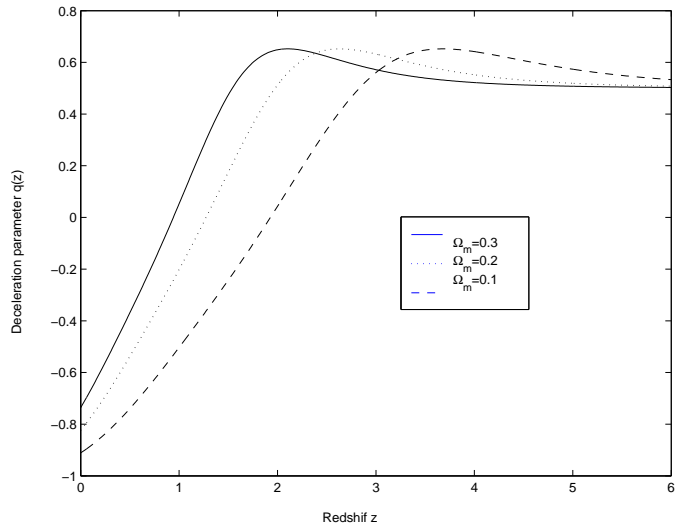}
  \caption{The dependence of deceleration parameter on redshift for $\Omega_m=0.1, 0.2, 0.3$ respectively. The solid,
  dotted, dash-dotted, dashed lines correspond to the full, first order, second order and third order MF
  equations respectively.}\label{1}
\end{figure}

\textbf{5. The Effective Equation of State}

Although in the modified gravity theory there is no need of
introducing dark energy. Following the general framework of Linder
and Jenkins \cite{Linder}, we can define the effective Equation of
State (EOS) of dark energy. Written in this form, it also has the
advantage that many parametrization work that has been done for
EOS of dark energy can be compared with the modified gravity.

Now following Linder and Jenkins, the additional term in Modified
Friedmann equation really just describes our ignorance concerning
the physical mechanism leading to the observed effect of
acceleration. Let us take a empirical approach, we just write the
Friedmann equation formerly as
\begin{equation}
H^2/H_0^2=\Omega _m(1+z)^3+\delta H^2/H_0^2\label{4.1}
\end{equation}
where we now encapsulate any modification to the standard
Friedmann equation in the last term, regardless of its nature.

Define the effective EOS $\omega _{eff}(z)$ as
\begin{equation}
\omega _{eff}(z)=-1+\frac{1}{3}\frac{d\ln\delta
H^2}{d\ln(1+z)}\label{4.2}
\end{equation}

Then in terms of $\omega _{eff}$, equation (\ref{4.1}) can be
written as
\begin{equation}
H^2/H_0^2=\Omega
_m(1+z)^3+(1-\Omega_m)e^{3\int_0^z{d\ln(1+z')[1+\omega
_{eff}(z')]}}\label{4.3}
\end{equation}

\begin{figure}
  \includegraphics[width=0.4\columnwidth]{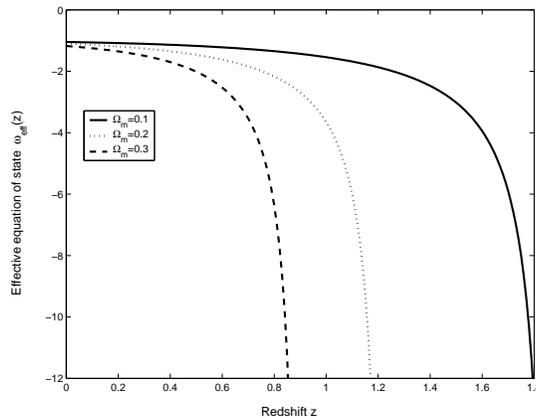}
  \caption{The dependence of effective equation of state on redshift for $\Omega_m=0.1, 0.2, 0.3$ respectively.}\label{1}
\end{figure}

Fig.5 shows the effective EOS computed with the MF equation
(\ref{MF}). We can see that it is consistent for all three values
of $\Omega_m$ with current bound: $-1.45<\omega_{DE}<-0.74$ (95\%
C.L.), \cite{Melchiorri}. A special property of the full EOSs are
that it would undergo a divergence at roughly $z=1.9, 1.25, 0.9$
for $\Omega_m=0.1,0.2,0.3$ respectively. After the divergence
point, the EOSs reduce to zero (We have verified this numerically,
but have not drawn them in Fig.5). What do these divergences imply
deserve further studies: maybe it is a catastrophe of this
modified gravity theory; maybe it merely indicate that the
definition of effective EOS is not suitable here: we just can not
consider the effects of modified MF equation as an equivalent
effects of dark energy.

\textbf{6. Discussions and Conclusions.}

In this paper we derived the full Modified Friedmann equation of
$1/R$ gravity in Palatini formulation. Although we only considered
the $1/R$ gravity, our derivation is actually quite general and
can be easily applied to other models of modified gravity of the
$L(R)$ type. For example, the modifications by adding $1/R+R^2$
\cite{Odintsov-R2} or $\ln R$ terms \cite{Odintsov-lnR} are also
suggested recently. Those two models may explain both the current
cosmic acceleration and early inflation without dark energy and
inflaton. Also quite interestingly, in those two models, the
instability of the metric formulation of the $1/R$ gravity may be
resolved. The Palatini formulation of those two models have been
studied in Ref.\cite{Wang2} using the framework in this paper.

We have also shown that the Palatini formulation of $1/R$ gravity
may accommodate the observational data indicating that our
universe expansion is accelerating without introducing dark
energy. On the other hand, the CMBR anisotropy recently measured
by WMAP gives another important data set. Any reasonable
cosmological model should fit those data. Whether the MF equation
derived in this paper can pass this test is our work ongoing.
Also, it is important to find whether the $1/R$ gravity can give
distinctive predictions from other models intending to explain the
cosmic acceleration such as cosmological constant, quintessence,
Chaplygin gas, etc. This is possible following the general
framework of Lue et al \cite{Lue}. There, the authors showed that
mechanisms of acceleration due to dark energy or Modified
Friedmann equation will give different predictions on the large
scale structure power specta and ISW effects. Applying their
framework to the MF equation (\ref{MF}) is straightforward. This
is also an interesting future working direction.

Whether the current observed cosmic acceleration is an indication
that gravity should be modified at large scale or there exist some
form of dark energy is surely one of the most important problem in
modern physics. Both of those two possibilities deserve be
investigated further.

\textbf{Acknowledgements}

We have benefitted a lot by helpful discussions with
R.Branderberger, A.Lue, D.Lyth, A.Mazumdar, S.Nojiri, S.D.
Odintsov, L.Ryder, X.P.Wu, K.Yamamoto and Y. Zhang.  This work is
partly supported by China NSF and Doctoral Foundation of National
Education Ministry.

\begin{appendix}
\end{appendix}


\begin{thebibliography}{99}
\bibitem{Perlmutter} S. Perlmutter el al. Nature \textbf{404} (2000) 955;
Astroph. J. \textbf{517} (1999) 565; A. Riess et al. Astroph. J.
\textbf{116} (1998) 1009; ibid. \textbf{560} (2001) 49; Y. Wang,
Astroph. J. \textbf{536} (2000) 531;
\bibitem{Spergel} D.N.Spergel, et al., Astrophys.J.Suppl. \textbf{148} (2003) 1 [astro-ph/0302207]
; L.Page et al. astro-ph/0302220; M.Nolta, et al,
astro-ph/0305097; C.Bennett, et al, Astrophys.J.Suppl.
\textbf{148} (2003) 175 [astro-ph/0302209];
\bibitem{hui} B.Roukema, et al., A \& A \textbf{382} (2002)397: A.Lidz, et al., astro-ph/0309204 and references therein;
\bibitem{Peebles} P. J. E. Peebles, B. Ratra, astro-ph/0207347; S.M.Carroll, Living Rev. Rel.
\textbf{4} (2001) 1 [astro-ph/0004075]; T. Padmanabhan, Phys.
Rept. \textbf{380} (2003) 235 [hep-th/0212290];
\bibitem{Lue} A. Lue, R. Scoccimarro and G. Starkman,
astro-ph/0307034;
\bibitem{Dvali} G. Dvali, G. Gabadadze and M. Porrati, Phys. Lett.
\textbf{B485} (2000) 208;
\bibitem{Carroll} S. M. Carroll, V.Duvvuri, M.Trodden and M.
Turner, astro-ph/0306438; S. Capozziello, S. Carloni and A.
Troisi, "Recent Research Developments in Astronomy \&
Astrophysics" -RSP/AA/21-2003 [astro-ph/0303041]; S.Capozziello,
Int.J.Mod.Phys.D \textbf{11} (2002) 483;
\bibitem{Odintsov-string} S. Nojiri and S. D. Odintsov, hep-th/0307071;
\bibitem{Chiba} T.Chiba, Phys.Lett. \textbf{B575} (2003) 1
[astro-ph/0307338];
\bibitem{Dick} R.Dick, gr-qc/0307052;
\bibitem{Dolgov} A. D. Dolgov and M. Kawasaki, Phys.Lett. \textbf{B573} (2003) 1 [astro-ph/0307285];
\bibitem{Vollick} D. N. Vollick, Phys.Rev. \textbf{D68} (2003) 063510 [astro-ph/0306630];
\bibitem{Meng} X.H.Meng and P.Wang, Class. and Quant. Grav. \textbf{20} (2003) 4949 [astro-ph/0307354];
\bibitem{Flanagan} \'{E}.\'{E}.Flanagan, astro-ph/0308111; ibid,
gr-qc/0309015;
\bibitem{Odintsov-R2} S.Nojiri and S.D.Odintsov, hep-th/0307288;
\bibitem{Odintsov-lnR} S.Nojiri and S.D.Odintsov, hep-th/0308176;
\bibitem{Wang2} X.H.Meng and P.Wang, astro-ph/0308284; ibid,
hep-th/0309062;
\bibitem{Krauss} L.M.Krauss and B.Chaboyer, astro-ph/0111597;
\bibitem{Fabris} J. C. Fabris, S. V. B. Goncalves and P. E. de Souza,
astro-ph/0207430;
\bibitem{Linder} E. V. Linder and A. Jenkins, astro-ph/0305286;
\bibitem{Melchiorri} A. Melchiorri, L. Mersini, C.J. Odmann and M. Trodden,
astro-ph/0211522;


\end{thebibliography}
\end{document}